\documentclass[12pt,journal]{IEEEtran}
\usepackage{graphicx}
\usepackage{url}
\usepackage{amsmath}
\usepackage{amssymb}
\usepackage{booktabs}
\usepackage{multirow}
\usepackage{hyperref}
\usepackage[english]{babel}
\usepackage[style = ieee]{biblatex}
\usepackage{enumitem} 
\usepackage{enumitem}
\usepackage{soul,color}
\usepackage{adjustbox}
\usepackage{comment}
\usepackage{csquotes}

\setlist[itemize]{align=parleft,left=0pt..1em}
\begin{document}
 \setlist{nosep}
\title{Envisioning the Future Role of 3D Wireless Networks in Preventing and Managing Disasters and Emergency Situations}
\author{Ahmed Alhammadi,~\IEEEmembership{Member, IEEE,}
Anuj Abraham,~\IEEEmembership{Member, IEEE,}\\
Aymen Fakhreddine,~\IEEEmembership{Member, IEEE,}
Yu Tian,~\IEEEmembership{Member, IEEE,}
Jun Du,~\IEEEmembership{Senior Member, IEEE,}
and Faouzi Bader,~\IEEEmembership{Senior Member,~IEEE}

\thanks{Ahmed Alhammadi, Anuj Abraham, Aymen Fakhreddine, Yu Tian, and Faouzi Bader are with the Technology Innovation Institute, 9639 Masdar City, Abu Dhabi, UAE. (e-mails: ahmed.alhammadi@tii.ae, anuj.abraham@tii.ae, aymen.fakhreddine@tii.ae, yu.tian@tii.ae, carlos-faouzi.bader@tii.ae)}

\thanks{Jun Du is with the Department of Electronic Engineering, Tsinghua University, Beijing, China. (e-mail: jundu@tsinghua.edu.cn)}
\vspace{-1cm} }

\maketitle

\begin{abstract}

In an era marked by unprecedented climatic upheavals and evolving urban landscapes, the role of advanced communication networks in disaster prevention and management is becoming increasingly critical. This paper explores the transformative potential of 3D wireless networks, an innovative amalgamation of terrestrial, aerial, and satellite technologies, in enhancing disaster response mechanisms. We delve into a myriad of use cases, ranging from large facility evacuations to wildfire management, underscoring the versatility of these networks in ensuring timely communication, real-time situational awareness, and efficient resource allocation during crises. We also present an overview of cutting-edge prototypes, highlighting the practical feasibility and operational efficacy of 3D wireless networks in real-world scenarios. Simultaneously, we acknowledge the challenges posed by aspects such as cybersecurity, cross-border coordination, and physical layer technological hurdles, and propose future directions for research and development in this domain. 
\end{abstract}

\begin{IEEEkeywords}
Three-Dimensional (3D) Wireless Networks, Disaster Prevention, Disaster Management, Emergency Situations, Vehicular Technology
\end{IEEEkeywords}

\section{Introduction}
\subsection{Overview}
Three-Dimensional (3D) wireless networks, characterized by the incorporation and synchronization of ground and aerial resources, have created a groundbreaking paradigm shift in wireless communication. Built to cater to diversified vertical communication needs, these networks hold significant promise to revolutionize disaster prevention and management procedures.
Building upon the paradigm shift initiated by 3D wireless networks, this work delves deeper into their potential applications and advancements. These networks, which harmonize terrestrial, satellite, and drone communication systems, are designed to meet a variety of communication needs. Their resilience to ground-level disruptions ensures uninterrupted communication, a critical factor in emergency scenarios.

This work provides insights into recent research focused on the integration of Satellite, Unmanned Aerial Vehicles (UAVs), and Terrestrial Networks (TN). These studies are integral in shaping the future of 3D wireless networks and their potential role in disaster prevention and management.

Transitioning from theoretical perspectives, the work explores the practical applications of 3D wireless networks. A plethora of use cases are presented, encompassing early warning systems, infrastructure monitoring, risk assessment, efficient evacuation procedures, and wildfire management. In the realm of wildfire management, the real-time, wide-area monitoring and alerts offered by these networks can expedite firefighting operations.

The real-world prototypes of these networks are also discussed, providing a comprehensive understanding of the challenges in the field, the current state of 3D wireless networks in disaster response, and their future potential. The work emphasizes the need for ongoing research and development for these networks to reach their full potential. It also spotlights the role of emerging technologies such as Artificial Intelligence (AI) in enhancing the performance and adaptability of these networks in various disaster scenarios.

\subsection{Motivation}
Disaster situations often disrupt and overstrain conventional communication networks due to a sudden surge in demand. This disruption can significantly slow down and impact the effectiveness of disaster response. Thus, we turn our attention to 3D wireless networks. These networks, known for their reliability and robustness, can be crucial during such crises.

\begin{figure*}[t]
	\centering
\includegraphics[width=.90\linewidth]{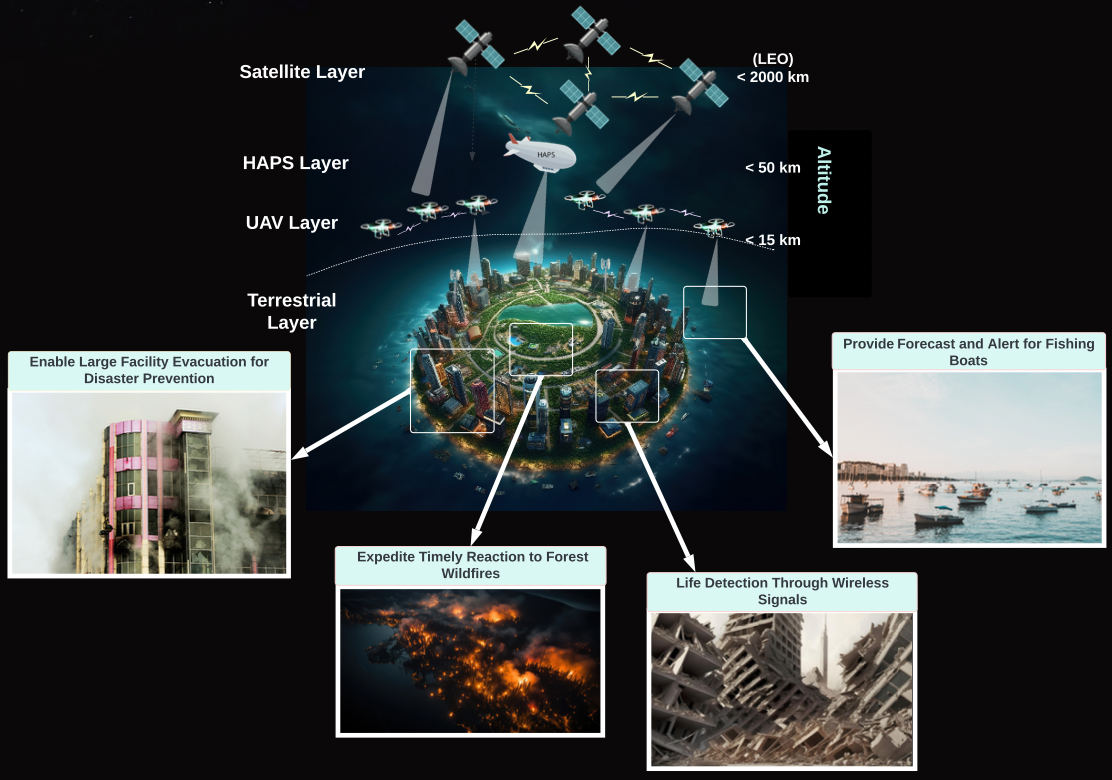}
	\caption{System Overview and Example Use Cases.}		\label{fig_1}
\end{figure*}

3D wireless networks combine terrestrial, satellite, and drone communication systems to maintain communication even during disasters. Figure \ref{fig_1} illustrates the system overview and example use cases in preventing and managing disasters. Unaffected by ground-level disruptions, these networks allow real-time data collection and transmission, improving situational awareness and aiding informed decision-making during emergencies.

Further motivation comes from advancements in vehicular technologies and the Internet of Things (IoT). These improvements offer the potential to improve evacuation procedures, coordinate rescue measures, and deliver essential services. Thus, 3D wireless networks become invaluable for efficient disaster management.

At its core, this exploration is fueled by the game-changing potential of 3D wireless networks in disaster response. Understanding and adopting these networks can lead us towards a future with disaster-resilient communication systems.
We summarize the main contributions as follows:
\begin{itemize}
    \item We provide a comprehensive review of the latest technological advancements in 3D wireless networks, focusing on their application in the field of disaster prevention and management. This review encompasses extensive literature on the integration of terrestrial, aerial, and satellite communications to form a unified system for crisis scenarios.
    \item Our work presents a novel assessment of various use cases where 3D wireless networks can play a transformative role, such as in early warning systems, infrastructure monitoring, risk assessment, and efficient large-scale evacuations. We offer insights into how these networks can significantly improve situational awareness and response times.
    \item We introduce pioneering solutions for seamless and rapid coordination during disaster management, demonstrating how 3D wireless networks can facilitate optimized routing, real-time communications, and swift deployment of emergency services. This includes an analysis of emergent technologies such as UAV-enabled Wi-Fi localization and AI-driven optimization.
    \item Through an exploration of several real-world prototypes, we illustrate the practical realization of theoretical models. These prototypes evidence the feasibility of employing 3D wireless networks in real-life disaster scenarios and highlight the potential for scaling up such technologies.
    \item We acknowledge and critically evaluate the current challenges faced by 3D wireless networking, including cybersecurity, regulatory cooperation, and physical layer issues. Subsequently, we propose strategic future directions for research and development to overcome these obstacles and leverage full network capabilities.
\end{itemize}
The rest of this paper is organized as follows. In Section~\ref{sec:state_of_the_art}, we introduce the foundational concepts and state-of-the-art developments in 3D wireless networks, delving into their architecture and the various components that constitute these systems. Moving on, Section~\ref{sec:disaster_prevention_management} provides a detailed exposition of the use cases for disaster prevention and management, drawing attention to the application-specific advantages and challenges. In Section~\ref{sec:real_world_prototypes}, we discuss real-world prototypes that have been developed, expounding on their design, functionality, and implications for disaster management. Section~\ref{sec:ntn_6g_mesh} focuses on the intersection of 3rd Generation Partnership Project (3GPP) Non-Terrestrial Networks (NTN), emerging sixth-Generation (6G) mesh networks, and how they relate to and enhance 3D wireless networks. Subsequently, Section~\ref{sec:ai_3d_wireless} analyzes the role of artificial intelligence within 3D wireless networks and its transformative effect across various facets of these systems. Following this, Section~\ref{sec:combined_use_case} presents a combined analysis of different use cases, accentuating the synergies between terrestrial, UAV, and satellite communications within 3D networks. Finally, Section~\ref{sec:challenges_future_directions} identifies prevailing challenges and potential avenues for future exploration, charting the course for progression in this domain. The paper concludes with Section~\ref{sec:conclusion}, summarizing the critical findings and prospects for the application of 3D wireless networks in disaster prevention and management.

\section{State-of-the-Art in 3D Wireless Networks}
\label{sec:state_of_the_art}

Recent years have seen significant progress in the development and deployment of 3D wireless networks. The growing body of research not only showcases technological advancements in this field but also highlights areas ripe for exploration.
\\
\\
\emph{Satellite-UAV-Terrestrial Integrated Network}:
\label{subsec:sutin}
The combination of Satellite, UAVs, and TN marks a significant advancement in the development of 3D wireless networks. This integration aims to enhance signal quality and reduce network latency. The study in \cite{mehdi} explores the pivotal role of UAVs as dynamic flying base stations in revolutionizing future communications networks by integrating with Device-to-Device (D2D) networks, thereby enhancing geographical coverage and optimizing data transmission rates. Building on the foundational insights provided by \cite{mehdi}, research by Nguyen et al. highlights the vital role of UAVs as flying relay stations, augmenting communication between satellites and ground stations in challenging scenarios \cite{nguyen2023real}. Donevski et al. evaluate in \cite{donevski2021experimental} the effectiveness of drone-mounted access points utilizing cellular infrastructure as backhaul in areas with mixed line of sight links. Experimental results show a coverage improvement by $6.4\%$ compared to classical cellular-direct links, with latencies consistently below 50 ms for $95\%$ of the experiment. Liu et al. further supported this integration by developing a joint subchannel assignment and power allocation algorithm. This optimizes the sum rate of secondary networks while addressing growing spectrum demands and imperfect Channel State Information (CSI)\cite{liu2023resource}.

Beyond enhancing network latency and signal quality, Yao et al. introduced an optimization scheme for robust and effective data acquisition in Internet of Remote Things (IoRT) systems. Their innovative approach leverages an integrated space-air-ground network to synchronize UAVs and Low-Earth Orbit (LEO) satellites for better operational efficiency \cite{yao2023optimization}.

In the domain of network security, Ren et al. assessed risks in UAV-aided satellite-terrestrial networks and proposed a Physically Unclonable Function (PUF)-based access authentication scheme. This novel solution ensures reliable mutual authentication, key agreement, and enhanced privacy protection in UAV operations \cite{ren2023novel}.

The innovative concept of integrating Intelligent Reflecting Surfaces (IRSs) into Satellite-UAV-Terrestrial (SUT) IoT networks was proposed by Liao et al. Their research demonstrated impressive network robustness against sophisticated jammers \cite{liao2023irs}. Ren et al. also addressed network security challenges and suggested a cutting-edge PUF-based access authentication scheme \cite{ren2023novel}. Furthering this line of thought, Chen et al. utilized the mobility of UAVs to optimize trajectories to serve a maximum number of users in Hybrid Satellite-Terrestrial Networks (HSTN) \cite{chen2022trajectory}.
These research examples underline the progressive strides made towards integrating Satellite, UAV, and TN. They offer a robust foundation for future advancements in these networks and their potential applications in disaster prevention and management.

\section{Disaster Prevention and Management}
\label{sec:disaster_prevention_management}
This section provides an overview of the use case for disaster prevention and management for 3D wireless network.

\subsection{Use Cases for Disaster Prevention}
Discuss the potential use cases of 3D wireless networks for disaster prevention, including early warning systems, infrastructure monitoring, and risk assessment.

\subsubsection{Large Facility Evacuation for Disaster Prevention}
Preventing disasters by a proactive approach (identifying hazards) can reduce the risk of potential damage and impacts of the disaster on society and the environment. A large facility evacuation system is necessary when the disaster is caused by either man-made (terrorists attack, traffic accidents, industrial explosions, fire hazards, or structure failures) or natural (floods, hurricanes, earthquakes, volcanoes, tornados, typhoons, or landslides). The information management system supports communication infrastructure such as, Wide Area Network (WAN)/Local Area Network (LAN), wireless LAN/WAN, satellite, General Packet Radio Service (GPRS), world wide web etc., to reduce disaster risks. 

The usage of efficient wireless cellular communications will make this framework a better way to communicate with the users to ensure/enable network resilience and redundancy. Through 3D wireless networks, sensors and UAVs or drones can be deployed to provide real-time information (camera images/videos) and then inspect the abnormal data from it with the use of AI technologies. Hence, this sensor information, satellite images, and with drones will help local governments to record the real-time situation of the specific location and provide a warning to the concerned authorities to act before the incident has happened.

Wireless signals encapsulate a wealth of intricate information, encompassing facets such as user mobility patterns and precise geospatial coordinates. The strategic harnessing of AI technologies for the comprehensive analysis and interpretation of user trajectories engenders a novel capacity: the identification and anticipation of imminent instances of congestion. Consequently, local governmental authorities stand empowered to orchestrate the deployment of pertinent law enforcement personnel and communication infrastructure, notably inclusive of mobile base stations. This strategic deployment engenders a framework for the efficient management, regulation, and mitigation of burgeoning congestion scenarios. This innovative paradigm exhibits scalability, seamlessly integrating across the spectrum of base stations, thus affording proactive vigilance over incidents of crowding, impervious to temporal fluctuations and adverse meteorological conditions, owing to the persistent interconnection upheld between mobile devices and the underlying base stations. This, in turn, engenders a symbiotic alliance with sophisticated sensory ensembles or UAVs, synergistically amplifying the real-time informational milieu concerning the intricate tapestry of crowd dynamics.

\subsubsection{Forecast and Alert Transmitter System for Cruise Ships, Merchant Marine and Vessels}
In the context of maritime safety, particularly for fishermen and cruise ship navigation, the integration of 3D wireless networks and other digital technologies can significantly enhance hazard forecasts and warning systems for cruises or fishing boats. This system provides a navigational aid to perform hazard forecasts and warning systems for fishing boats before turning into a disaster. Possibly, a transmitter sends different types of emergency signals for fire, boat/vessels or cruise ships sinking, man overboard, or medical help on manual activation. The system also extends support by providing navigational assistance for safe returns.

Moreover, the system leverages the capabilities of 3D wireless networks to provide localization and awareness of the fishing boat or vessels not crossing international territory, i.e., gives alarm to the fisherman onboard when his boat approaches the international border, and announces distance in his mother tongue, in his colloquial tongue. In addition, a Very High Frequency (VHF) based radio communication system for the fishermen's community has been established that sets up voice calls between the crafts and then sends the position data to the shore stations. This system can function as a ‘black box’ like one in aircraft and will help search operations in case of missing boats. Moreover, AI-based UAVs can provide a holistic view for situational awareness and helps in the rescue of missing fisherman \cite{OJCOMS2}.

\subsubsection{Timely Reaction to Wildfires}

In recent years, the world has witnessed an alarming surge in the frequency and scale of wildfires, posing significant challenges to both local and national authorities. To address this growing concern, we propose to leverage 3D networks, comprising ground-based sensors, drones, and satellite communication systems, to enhance early detection and timely management of large-scale of wildfires. This approach promises to strengthen our ability to combat these devastating natural disasters. In this section we examine how 3D networks can be beneficial for \textbf{\textit{i)}} early detection and \textbf{\textit{ii)}} large-scale wildfire management.

For accurate early detection of wildfires, we envision a three-layer 3D network that comprises a network of IoT sensors strategically placed in wildfire-prone areas and that send frequent short messages to a central control unit via Narrowband-IoT (NB-IoT) technology.  This real-time data transmission ensures swift detection of incipient wildfires, giving authorities crucial information in timely manner to swiftly organize operations and limit the damage. This is complemented by fixed wing UAVs that patrol vast areas at high altitudes to relay critical information to the control unit via satellite communication or direct laser or Free Space Optical (FSO) communication. This aerial surveillance provides extensive coverage, allowing for the early detection of wildfires across large territories. Furthermore, for a closer examination of high-risk areas, quadcopter UAVs or similar can easily be launched from areas near to forest can fly at low altitudes to keep an eye on vulnerable regions and report their findings to the control unit via cellular technologies or direct wireless links to the fixed wing UAVs if they are in the vicinity. Their ability to navigate close to the ground ensures a detailed assessment of potential fire threats.

After an expeditious detection of a wildfire, we propose to use 3D networks in a manner that facilitates harnessing large-scale wildfires. Drones can be valuable assets in responding to wildfires thanks to their 3D mobility, ability to fly at low altitude and record live videos that can be crucial in understanding the extent of the damage but also spot potential routes that can speed-up ground intervention, and fast and massive deployment which reduces the risk on endangering human operators. 

UAVs bring for instance added valuable capabilities in terms of live video streaming, thermal imaging, and high-resolution mapping technology allowing for fast delivery of real-time data to central crisis handling units, empowering decision-makers with comprehensive insights. This data aids in understanding the wildfire's scope, identifying optimal routes for ground intervention, and facilitating informed crisis management. 

Coordination of first responders could therefore benefit from the critical role played by UAVs in orchestrating the efforts of both ground-based and aerial first responders. In this case, the communication layers in place are at three main levels: Ground, aerial and satellite. The aerial layer can by itself be divided into three sub-layers, namely 
\begin{enumerate}
    \item low-altitude UAVs,
     \item helicopters,
     \item and high altitude fixed wing UAVs together with aerial firefighting jets
\end{enumerate}
The use of satellite here is of particular interest to send command messages to airborne firefighters and for cross-layer and sub-layer communication in case of a lack of direct communication link via laser or FSO. To sum up, our proposed system architecture outlines distinct communication layers for efficient and seamless communication allowing for both local coordination and control via ground and low-altitude UAVs as well as long-range communication and wide-area coordination by articulating high-altitude coordination and communication among aircrafts and satellites. 

\begin{figure*}[t]
	\centering
\includegraphics[width=.90\linewidth]{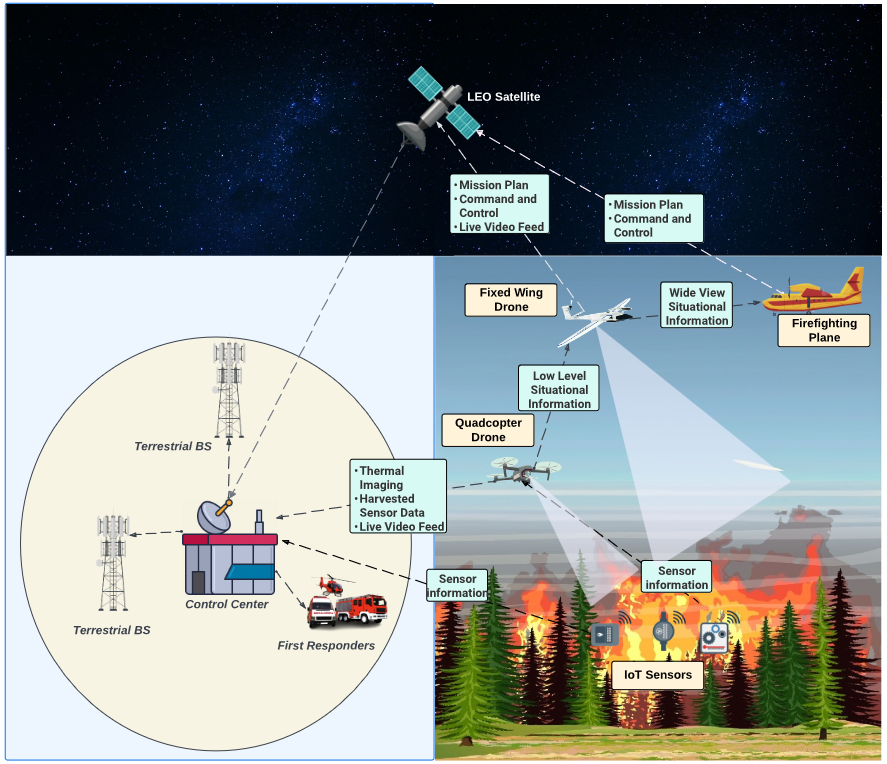}
	\caption{3D Network For Wildfire Detection and Management.}		\label{fig_2}
\end{figure*}
Building on this proposal, recent research further explores the potential of integrating emerging technologies for wildfire detection and management. The study \cite{F1} conducted by Ahmad et al. introduces FireXnet, a deep learning model designed specifically for efficient wildfire detection. The model, notable for its lightweight architecture, achieves high accuracy with significantly less training and testing time. By incorporating SHAP (SHapley Additive exPlanations), an explainable AI tool, the model's interpretability is enhanced, aiding in the identification of key characteristics triggering wildfire detections. The incorporation of such an AI system with 3D networks could result in a multi-layered early warning system, thereby improving the speed and accuracy of wildfire detection.

Further advancements are discussed in \cite{F2} by Suganya \textit{et al.}, which presents a forest fire detection system based on Wireless Sensor Networks (WSN). The system capitalizes on the ability of WSNs to detect changes in environmental parameters related to fire events. It then analyzes this data to distinguish between normal environmental fluctuations and fire-related anomalies. Additionally, the system employs a machine learning model that integrates neural prophet and logistic regression for forest fire prediction. The combination of real-time detection via WSN and future predictions through machine learning could significantly amplify the effectiveness of large-scale wildfire management strategies.

In a similar vein, the work in \cite{F3} evaluated the effectiveness and reliability of a WSN for detecting forest fires and providing timely alerts in its research. The authors conducted field experiments that simulated real-world fire conditions in a forested area, and the WSN demonstrated promising results. This finding underscores the potential effectiveness of using a wireless sensor network to enhance fire management strategies. It could lead to more timely responses, reduced response times, and minimized impact of forest fires on ecosystems and communities.

Finally, the system architecture proposed in the original paper, featuring distinct communication layers for efficient and seamless communication, is worth highlighting. This includes local coordination and control via ground and low-altitude UAVs, as well as long-range communication and wide-area coordination through the use of high-altitude coordination and communication among aircraft's and satellites. The use of different types of UAVs at various altitudes for surveillance, close examination, and large-scale wildfire management could offer a comprehensive and efficient solution for timely wildfire detection and management. An accompanying Fig. \ref{fig_2} likely offers a visual representation of this multi-layered communication system, illustrating how each type of aerial vehicle operates at different altitudes and contributes to the overall system.

\subsection{Use Cases for Disaster Management}
In this subsection, we examine the potential use cases of 3D wireless networks for disaster management, including real-time communications, rapid deployment of emergency services, and effective coordination of rescue efforts. Main use cases are hereafter mentioned:

\subsubsection{Large Facility Evacuation for Disaster Management}
Human-made disasters or natural disasters can disrupt the whole community including threats to people, property, economies, and the environment. Effective disaster management often hinges on the strategic organization of community resources, and the use of 3D wireless networks can greatly enhance these efforts, particularly in the context of large facility evacuations. Moreover, disaster relief management include various operations such as evacuation, search and rescue operations, and emergency medical assistance. Towards this, following a proper evacuation and relief response procedure with suitable efficient wireless communication in the proposed framework of the evacuation management system will support emergency preparedness and resilient reconstructions in a better way.

In the proposed evacuation framework, leverages 3D networks with sensors deployed at specified sites can collect all relevant information, such as hazardous area location, the number of households affected, and disaster intensity, to evaluate the level of the potential threat. After that, analysis based on the collected data will give the inputs for the next step, which includes the evacuee population, specified pick-up and shelter locations, the evacuation time window, and accessible routes in the surrounding road network.

In the next stage, the estimated inputs are used to generate the evacuation plan, which is mainly divided into three categories: demand management, supply management, and mitigation management. Both demand and supply management focus on the operation conducted in the areas affected or areas adjacent to the hazard zone. While mitigation management is more about diverting traffic and sending a warning at the peripheral area to prevent incoming vehicles and pedestrians from entering the hazard zone. Demand management solves the evacuee transfer problems such as shelter assignment, stage-based evacuee releasement, and vehicle routing to carry evacuees from hazard to zone to specified shelters. On the other hand, supply management focuses on providing convenience for transferring evacuees. The most common strategies include capacity reversibility (contraflow strategy) and adaptive signal control to provide either more capacity or more smooth accessibility for emergency vehicles. At the final stage, the corresponding emergency departments or agencies implement the obtained evacuation plans.
In summary, the integration of 3D wireless networks in our proposed evacuation management system enhances the strategic organization of resources, promotes efficient communication, and supports the effective execution of evacuation plans, ultimately contributing to more resilient disaster management.


\subsubsection{Group-based Emergency Transportation Routing for Disaster Response}

In the face of large-scale disasters, particularly in densely populated public facilities, traditional emergency transportation routing models including UAVs, helicopters, and others transportation engines for first care or primary emergency, may fall short in their coordination. The potential for a high number of victims in such scenarios necessitates a more comprehensive and efficient approach to providing immediate medical aid. Here, 3D wireless networks can significantly improve the coordination and assistance of emergency transportation routing by providing timely, efficient communication for a large geographical area.

3D wireless networks can  play an important role in coordinating and assisting the emergency vehicle routing to reach destination faster by enabling guaranteed wireless connectivity. Highly reliable connectivity combined with information from UAVs on situational awareness could make emergency transportation routes more intelligent, enabling emerging tools such as artificial intelligence to further improve real-time services.

Different from the emergency vehicle routing models for disaster response management used in household and on-road traffic accident situations, victims could be a large population when attacks happen in a densely populated public facility. Dispatching primary/first aid  transportation from a single specified hospital to the facility may not be sufficient to provide first aid for such huge demand. On this basis, emergency vehicle from more hospitals or even all hospitals in an entire city may need to be dispatched to rescue the injured people; therefore, a Community Covering Problem (CCP) will not exist in this situation. We shall dispatch emergency vehicle to multiple hospitals to cope with this problem and to efficiently utilize the emergency vehicle, the patients are divided into two groups, either slightly injured or seriously injured. Also, if the patient is slightly injured, the emergency transportation can assist people on the field without carrying them back to the hospital, then this emergency vehicle can visit the next patient, either slightly injured or seriously injured.

On the other hand, if the emergency transportation provides aid to seriously injured people, then it will carry the patient directly back to the hospital without any hesitation. The objective of the optimized routing problem is to minimize the longest waiting time of a patient in two groups, by doing so; we can make maximum use of the emergency transportation, and guarantee system optimality for all patients with different severity. In addition, the emergency transportation routing problem is no longer a single pick-up and drop-off problem.

\subsubsection{Life Detection through Wireless Signal}
The paradigm of life detection through wireless signals emerges as a pioneering avenue within the domain of disaster prevention and management. Leveraging the inherent characteristics of wireless transmissions, such as signal propagation and reflection, this innovative approach transcends traditional methods by enabling the detection of human presence and movements even in scenarios marked by obscured visibility or compromised physical access. Incorporating 3D wireless networks into this framework can further enhance the resiliency and reach of this innovative method. By employing advanced signal processing techniques and machine learning algorithms, the subtle perturbations induced by human activities upon wireless signals can be discerned, allowing for the identification of survivors, trapped individuals, or even potential hazards in disaster-stricken environments. This transformative capability not only enhances search and rescue operations conducted by ground personnel but also integrates seamlessly with emerging technologies like drones acting as flying base stations and satellite communications systems. These aerial platforms and satellites extend the reach of wireless signal detection to inaccessible or hazardous zones, facilitating real-time tracking of survivors and expediting the deployment of rescue teams. However, this approach does entail challenges such as signal noise, environmental interference, and ethical considerations, which necessitate careful exploration and mitigation. As the technological landscape continues to evolve, life detection through wireless signals, bolstered by the synergy with drones, flying base stations, and satellite communications, stands as a promising frontier. This approach promises to reshape the landscape of disaster management by offering a versatile, non-intrusive, and potentially life-saving methodology.

\subsubsection{IoT Enabled Global Route Planning \& Optimization for Emergency Buses}
Improving the positioning accuracy from meters to centimeters will enhance route planning for driving navigation and other high-precision user positioning services. Leveraging cellular-based IoT through 3D wireless networks can facilitate rapid evacuation for emergency buses during disaster management. This will provide optimal routing protocols, by considering the traffic information collected from the mobile base stations placed at the subsequent intersections and coordination among them via satellite. To ensure reliable, safe, and efficient communication for routing emergency buses on road networks is required to provide technological solutions in intelligent transportation systems.

This allows public buses to be used as emergency buses during the disaster relief process by communicating with the road infrastructures, pedestrians, etc so that optimal global routing of buses is established for evacuation. Here, 3D wireless networks help establish optimal global routing for evacuation purposes. In scenarios where fixed radio communication infrastructures or advanced sensors fail, UAVs can restore these tasks, ensuring uninterrupted communication.

Due to the large ridership and flexible driving routes, public buses are a good choice for transferring evacuees during evacuation. So, we assume the city has emergency bus terminals, specified emergency pick-up points, and shelters scattered around the whole city. The emergency pick-up points can be a subset of the bus stops, and their usual function is the same as the normal bus stop if no disaster happens, otherwise, evacuees will evacuate from the incident site towards their closest emergency pick-up stops. Each of these pick-up points is installed with advanced sensors to detect the population of gathered evacuees in real-time.

\subsubsection{Transit Signal Priority (TSP) in Emergency Evacuation}
Real-time traffic dispatch for rapid disaster response is necessary to provide new emergency communication insights by integrating the terrestrial layer with high-quality V2X (vehicle-to-everything) services and NTN through 3D wireless networks. This will improve precision positioning, navigation, and reduces dwell times of public bus priority or TSP when it reaches a signalized intersection during emergency evacuation.

V2X communication with the roadside unit and combining with IoT sensors gather information remotely via Wireless Local Area Network (WLAN), Dedicated Short-Range Communication (DSRC), cellular-based, or any other related technologies, is the standard way of linking vehicles, road infrastructures, pedestrians, etc., towards the emerging smart city concept. IoT communications are implemented based on 3D wireless networks.. During the disaster relief process, priority emergency vehicles have to move very easily on the road for safe pick-up and drop-off of evacuees.

\subsubsection{UAV-Enabled Wi-Fi Indoor Localization and Navigation for Multi-Floor Rescue and Evacuations}
In multi-floor emergency scenarios like building collapses or fires, indoor Wi-Fi devices are high likely damaged and the wireless signals from base station can't detect the heights or precise locations of the users. In this context, UAVs equipped with Wi-Fi transmitters and integrated with 3D wireless networks offer a groundbreaking solution. These UAVs serve as flying access points, establishing an adaptable Wi-Fi network for indoor localization and navigation. Imagine a collapsed building due to an earthquake; strategically positioned UAVs  act as Wi-Fi hubs. Victims and rescue teams carry Wi-Fi-enabled devices that connect to the UAV network. Cutting-edge algorithms interpret Wi-Fi signals, allowing precise positioning across floors. Victims receive real-time directions to safety, while responders navigate hazardous terrain using UAV-enabled guidance.

The advantages are significant: UAVs extend network coverage to inaccessible zones, aiding swift and efficient multi-floor rescue efforts. This technology enhances coordination among rescue teams, optimizing outcomes in high-pressure scenarios. Despite challenges like signal interference, the fusion of UAVs and Wi-Fi opens new avenues for successful multi-floor rescue and evacuations, elevating disaster response to unprecedented levels of effectiveness.

\section{Real-world Prototypes}
\label{sec:real_world_prototypes}
Developing and validating prototypes in real-world settings is crucial in the field of disaster management. Prototypes allow for testing and refining of concepts, technologies and strategies before they are implemented in real-life scenarios where lives and resources are at stake. They provide valuable feedback and data, that helps in identifying the potential pitfalls and areas of improvement, which might not be evident in theoretical models or simulations. Moreover, real-world prototypes facilitate the assessment of operational feasibility, effectiveness and efficiency of proposed solutions. They also facilitate iterative development, where each version of the prototype is refined based on the observations and data gathered from its preceding version. This leads to a well-tested and reliable solution that can be deployed in actual disaster scenarios. Table \ref{tab:realworldprototypes} summarizes three real-world prototypes that are discussed in the following subsections.

\begin{figure}[t]
\centering
\includegraphics[width=\linewidth]{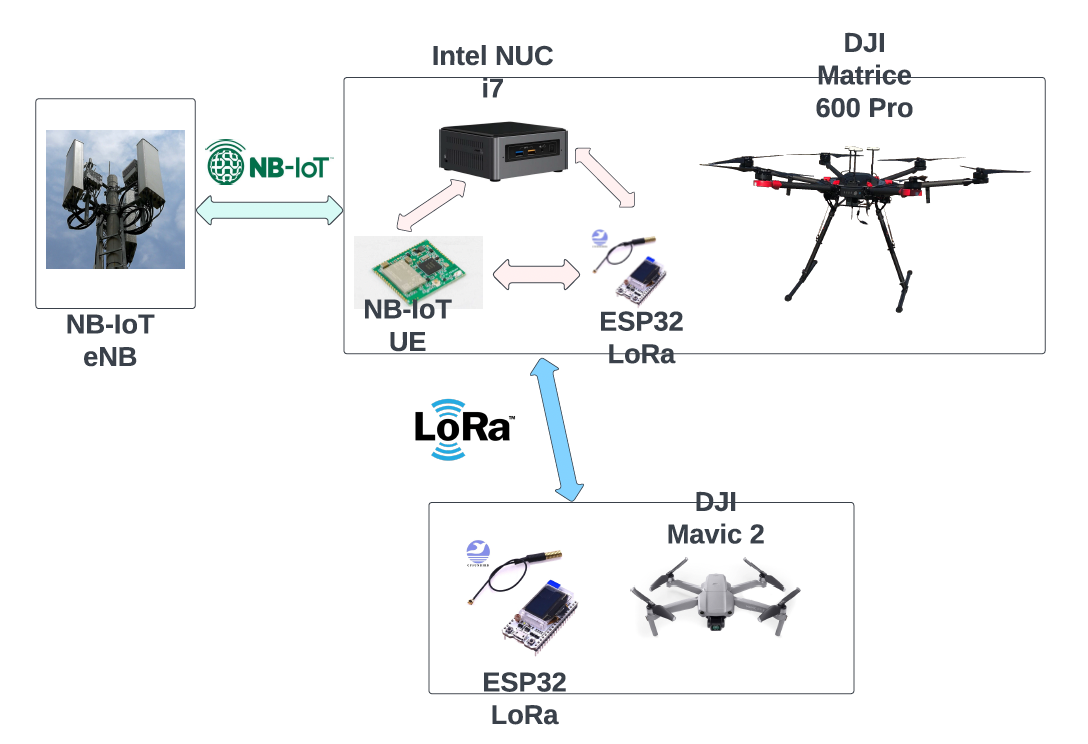}
\caption{Tier 1 NB-IoT and Tier 2 LoRa System for Prototype 1 \cite{P1}. \label{fig:fig_3}}
\end{figure}

\subsection{Prototype 1: Two-Tier UAV-based Low Power Wide Area Networks}
The prototype, titled ``Two-Tier UAV-based Low Power Wide Area Networks: A Testbed and Experimentation Study as shown in Fig. \ref{fig:fig_3}, integrates cutting-edge Low Power Wide Area Network (LPWAN) technologies with UAV systems to address connectivity challenges in remote rural environments \cite{P1}. This innovative two-tier network architecture employs UAVs as mobile base stations in areas not covered by traditional infrastructure. The first tier of this network is the ground-based LPWAN deployment, where both NB-IoT (Narrow Band Internet of Things) and LoRa (Long Range) stand as potential technologies. NB-IoT, with its cellular-based approach, leverages existing 3D wireless cellular networks for secure and reliable communications, optimized for a large number of devices and indoor penetration \cite{arum}. Conversely, LoRa, with its long-range capabilities and operation on unlicensed frequencies, offers a cost-effective, power-efficient solution ideal for less dense device deployments. When Tier 1 LPWAN coverage is inadequate, the second tier, comprising UAVs, can be rapidly deployed to ensure continuous connectivity for critical IoT applications such as precision agriculture, forest management, and wildlife monitoring, thus overcoming the limitations of TN and bringing the benefits of both NB-IoT and LoRa technologies to the most inaccessible areas.

This prototype has been successfully deployed and utilized in an actual rural environment, demonstrating the two-tier LPWAN network's effectiveness in extending network coverage where traditional macro-cellular networks cannot reach. The mobile Tier 2 LPWAN base stations, mounted on UAVs, provide connectivity to static or mobile LPWAN user equipment in areas that lack direct Tier 1 LPWAN network coverage.

\subsection{Prototype 2: An Intent-Based Reasoning System for Drone Missions}
\begin{figure}[t]
	\centering
\includegraphics[width=.90\linewidth]{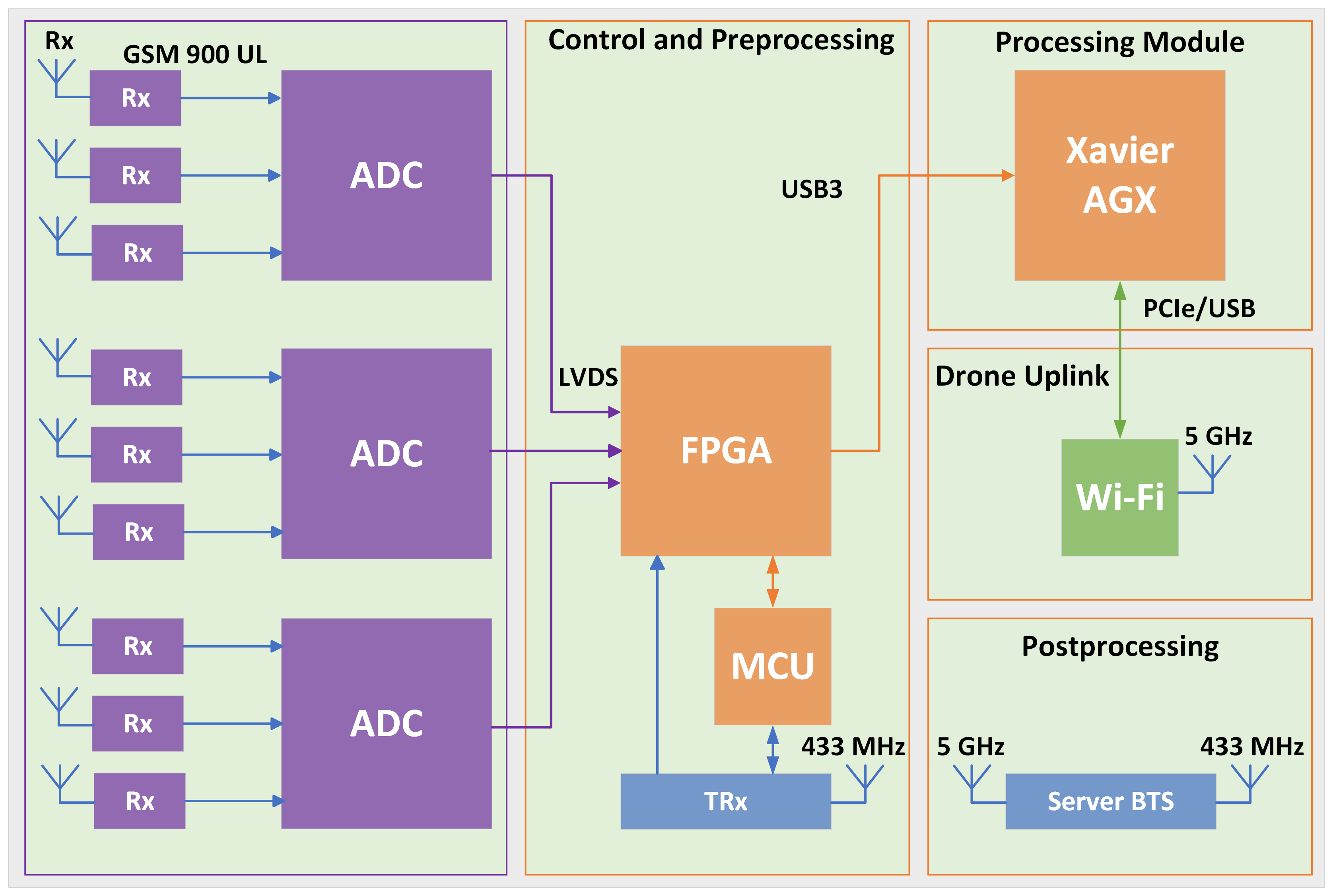}
	\caption{A block diagram of the UAV-based phone localization system used in Prototype 2 \cite{P2}. \label{fig:fig_4}}
\end{figure}

This prototype aims to automate the generation of drone missions based on user intent as depicted in Fig. \ref{fig:fig_4}. The system utilizes machine learning models to understand user intent expressed through spoken language and then translates this intent into mission parameters for drones. This could be of significant use in search and rescue operations, where rapid response and efficient resource allocation are critical.
This prototype \cite{P2} was demonstrated in both urban and rural environments. The results showed the feasibility of using intent-based reasoning systems for controlling autonomous drones in search and rescue operations. The system could reliably capture user's intent, translate it into drone mission parameters, and then control the drone to execute these missions. This shows the potential of combining AI technologies with drone technology to enhance disaster management operations.

\subsection{Prototype 3: A Direction-of-Arrival Estimation System for UAV-Assisted Search and Rescue}
\begin{figure}[t]
	\centering
\includegraphics[width=.90\linewidth]{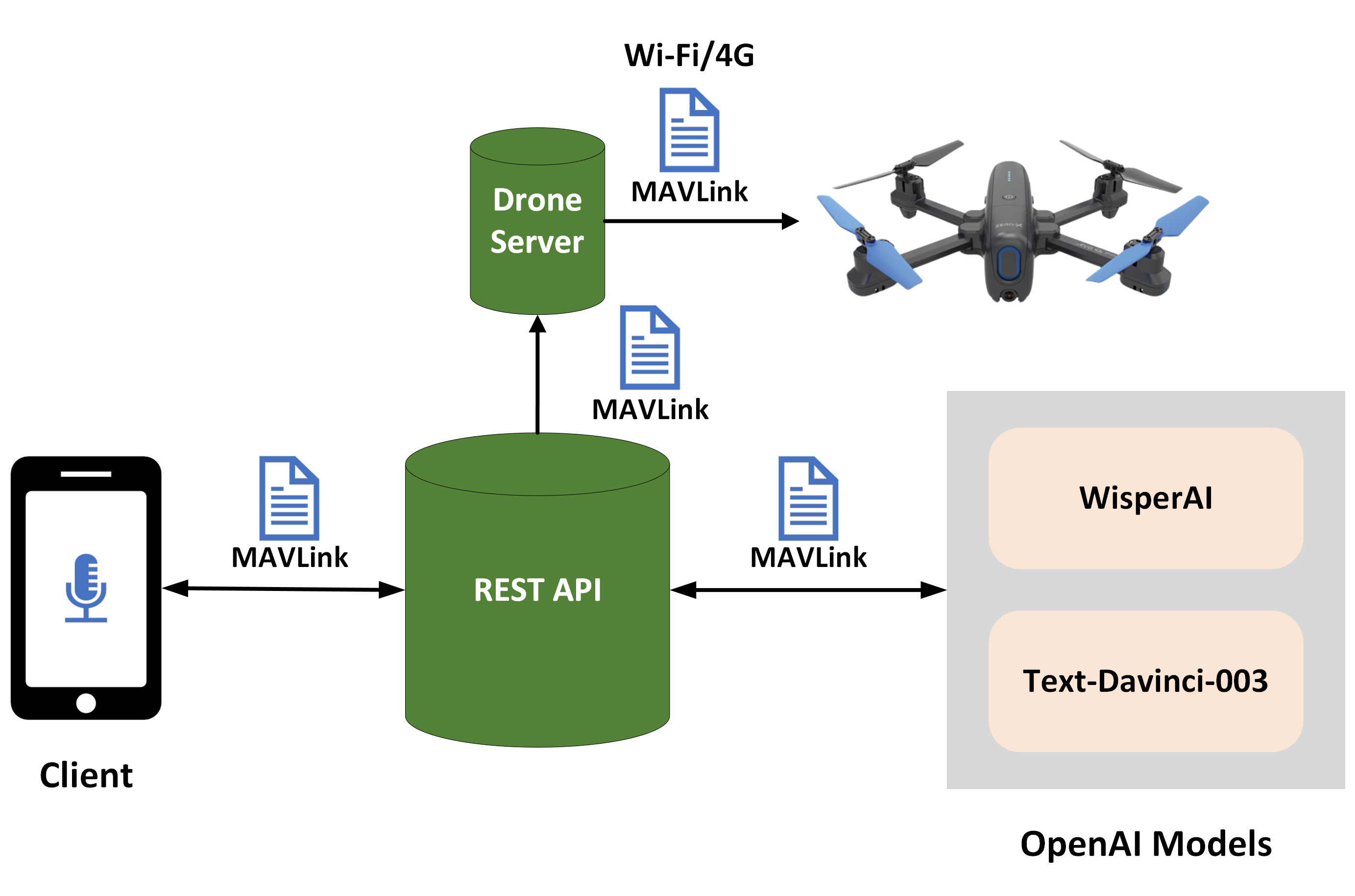}
	\caption{System Overview for Prototype 3 \cite{P3}.  \label{fig:fig_5}}
\end{figure}
The prototype shown in Fig. \ref{fig:fig_5} is a drone-based multi-functional system, SORTIE \cite{P3}, designed for Search and Rescue (SAR) operations. The system aims to supplement traditional SAR methods by using mobile phones as indicators of buried victims in disaster-stricken areas. The drone, equipped with a subsystem, can be launched shortly after SAR teams arrive to map and photograph the affected area. Upon returning, a phone localization module is attached for the next flyover, during which it collects data for Direction-of-Arrival (DOA) estimation. This data is then processed on a Ground Station (GS) to estimate the most promising locations for the SAR team to search. The system's operation does not interfere with the ongoing safety assessment process of the rescue operation.

The system is based on the Global System for Mobile Communication (GSM), chosen due to its universal support by virtually all mobile phones. The Extended GSM (E-GSM) 900 band was selected for its relatively low frequency and lower attenuation through building materials, ensuring a high Signal-to-Noise Ratio (SNR) conducive for reception through rubble. The system includes a GSM Base Transceiver Station (BTS), a smart jammer to disrupt available mobile phone networks, and a server for data post-processing and visualization. The BTS controls the phone's exact transmission time and frequency. The airborne component of the system consists of an antenna array, multichannel phase-coherent receivers, synchronization hardware for the BTS, and a general-purpose processing system. The system's weight and power consumption are critical factors, and its design aims to be lightweight and energy-efficient. The paper acknowledges the challenges of radio propagation in rubble piles and employs diversity and advanced algorithms to overcome these issues.

\subsection{The Need for A Fully 3D Satcom-UAV-Terrestrial Prototype}

Despite the significant strides made in the development of UAV-based communication systems for disaster management, there is still a conspicuous gap in this field. A fully-integrated, 3D Satellite - UAV - terrestrial prototype is yet to be developed. This type of prototype would integrate ground-based systems, aerial systems (UAVs), and satellite systems into a comprehensive, multi-layered communication network. A 3D Satcom-UAV-Terrestrial prototype would provide several advantages over current systems that we mention hereafter:
\begin{itemize}
    \item It would offer a truly comprehensive coverage, combining the wide coverage of satellite systems, the flexibility and agility of UAVs, and the accessibility of terrestrial systems. This would ensure continuous and reliable communication in all areas, even in the most remote or disaster-stricken areas where ground infrastructure might be damaged.
    \item Such a prototype would enhance the robustness and resilience of the communication network. By integrating multiple layers and types of communication systems, the network would be able to continue functioning even if one layer fails or is compromised. This is particularly important in disaster scenarios, where infrastructure might be destroyed or inaccessible.
    \item A 3D Satcom-UAV-Terrestrial prototype would facilitate efficient resource allocation and coordination. By providing a comprehensive view of the entire operation area, it would allow for real-time monitoring and control of resources, enhancing the effectiveness of disaster response efforts.
\end{itemize}

The significant potential benefits of a fully 3D Satcom-UAV-Terrestrial prototype make it crucial to address existing challenges and develop such a prototype. This requires the combined efforts of researchers, engineers, policymakers, and disaster management stakeholders. In the next section, we will further analyze the added value of a 3D wireless network in relation to each of the use cases mentioned.

\section{3GPP Non-Terrestrial Network (NTN) Standards and 6G Mesh Networks}
\label{sec:ntn_6g_mesh}
This section briefs the recent developments in the integration of NTN systems with 5G networks in 3GPP standards and explains the 6G mesh resilience network technology to support NR-based satellite access and D2D services.
 
\subsection{NTN in 3GPP Standards}

\begin{figure}[t]
\centering
\includegraphics[width=.95\linewidth]{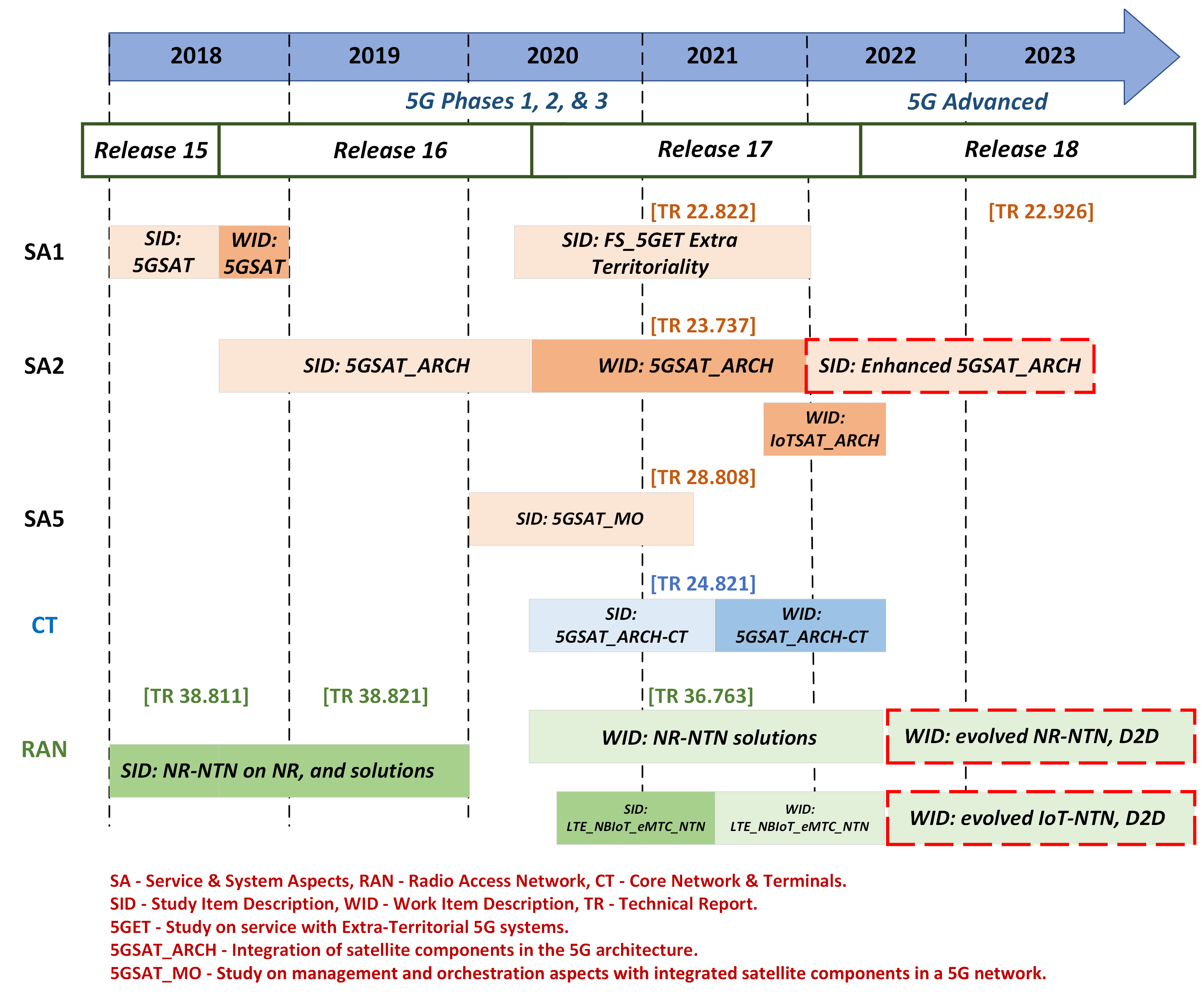}
\caption{Current Roadmap of Different Activities in 3GPP NTN.} \label{fig_NTN}
\end{figure}

Figure \ref{fig_NTN} shows the current roadmap of different activities carried out in 3GPP related to NTN systems toward a global standardized solution. It started in Release 15 and Release 16 with some initial studies and work item descriptions and their dependencies. In Release 15, a Technical Study (TS) on New Radio (NR) to support NTNs [TR 38.811] and enhanced LTE support for aerial vehicles [TR 36.777] was carried out. Whereas, in Release 16, a Technical Report (TR) on the solutions for NR to support NTNs [TR 38.821] and remote identification of UAVs [TS 22.825] for the potential requirements and use cases were reported~\cite{3GPP}.
Later, Release 17 focuses on transparent payload architecture with a frequency band below 6 GHz and covers both earth-moving and earth-fixed radio cells [TR 23.737], IoT and 5G access via satellite non-terrestrial link~\cite{release17}.
In addition, NB-IoT and enhanced Machine-type Communication (eMTC) support for NTN [TR 36.763] was also studied with 5G enhancements for UAVs [TS 22.125, TS 22.829] \cite{white}. It also supports \cite{rel17}:

\begin{itemize}
    \item Addressing identified issues due to long propagation delays, large Doppler effect, and moving cells in NTN systems.
    \item NR-NTN based on LEO and GEO with implicit compatibility to support HAPS and ATG (Air to Ground) scenarios.
    \item Transparent NG-RAN (Next Generation Radio Access Network) architecture enhancement and related procedures.
    \item Service continuity from TN to NTN and from NTN to TN systems, and
    \item eMTC-based satellite access to address massive IoT.
\end{itemize}

As a part of Release 18, several features have been proposed. Some of these include but are not limited to network-based trusted User Equipment (UE) location determination, TN/NTN service continuity, relay-based architecture including satellite, coverage enhancement, NR-NTN deployment in above 10 $GHz$ bands, network-based UE location, sidelink/D2D services, NTN-TN and NTN-NTN mobility and service continuity enhancements~\cite{release18}. On the IoT, one important feature to be considered is the store and forward (regenerative payload for Cellular-IoT). Other IoT services and features include discontinuous coverage, mobility enhancements, etc. Moreover, the possible new RAN features include asynchronous multi-connectivity and carrier aggregation, coordinated transmission, UE without Global Navigation Satellite System (GNSS) capabilities, NTN-TN spectrum coexistence, and regenerative payload with edge computation.

Moving toward the era of 6G deployment, a glimpse of 6G NTN systems targeted the following features, such as constellation available capacity of 12 $Pbps$, each satellite capacity of 1.5 $Tbps$, a coverage area of approximately 600,000 $km^2$, the peak throughput and per beam throughput of 18 $Gbps/km^2$ and 50 $Mbps/km^2$ respectively. The proposed 6G NTN services include global extreme low latency links, 40 $Mbps$ to devices, autonomous driving for global coverage, remote sensing and precision positions, and new digital architecture for un-connected regions. 6G NTN for low latency should ensure a long-distance global low-latency service. Similarly, 6G NTN for Ultra-Reliable Low Latency Communications (URLLC) requires low latency and high-reliability bent pipe (approx 2 $ms$), especially where there is no coverage, and MEC (Mobile Edge Computing) to support wide-area real-time control.

\subsection{Emerging 6G Mesh \& Resilient Network Technology}

The progression toward 6G heralds a new era for mesh networks, characterized by their decentralized nature and enhanced resilience. In the context of disaster prevention and management, such networks' ability to dynamically reroute traffic and maintain connectivity despite individual node failures is invaluable.

\paragraph*{Manpack Base Stations}
A novel approach to ensuring network resilience in 6G mesh networks is the deployment of manpack base stations \cite{concha2023harnessing}. These portable units, carried by individuals, can quickly establish mobile network nodes in areas where traditional infrastructure is compromised or nonexistent. Manpack base stations extend the network's edge, providing critical connectivity for emergency responders and affected individuals in disaster-stricken zones.

\paragraph*{Network Slicing}
Network slicing in 6G mesh networks offers the ability to create isolated virtual networks tailored to specific services or operational requirements. This technique can be employed to prioritize emergency communications, ensuring that critical messages and data are transmitted with precedence during disasters. Network slicing also allows for the efficient allocation of resources, adapting to the fluctuating demands of a crisis without compromising other network services \cite{rodrigues2022network}

\paragraph*{D2D and V2X Communications}
D2D and V2X communications are essential in maintaining a resilient mesh network \cite{he2021d2d,}, enabling direct communication pathways between devices and vehicles. This capability is particularly significant when infrastructure-based connectivity is compromised. In a 6G mesh network, D2D and V2X can uphold a communication fabric among users, emergency vehicles, and service drones, ensuring information flow remains uninterrupted in the face of infrastructure damage.

\paragraph*{Self-Organizing Networks (SON)}
SON \cite{papidas2022self} capabilities will be integral to 6G mesh networks, allowing for autonomous configuration, optimization, and healing. Such self-management features are vital for disaster scenarios, enabling the network to adapt to node losses or surges in demand without human intervention. The incorporation of SON into 6G mesh networks ensures a level of operational continuity that is essential for effective disaster response.

By harnessing these advanced technologies and techniques, 6G mesh networks will not only maintain communication during disasters but also enhance situational awareness and coordination of emergency services. As the development of 6G continues, further research into these enabling technologies will solidify the foundations of a truly resilient and adaptive communication infrastructure for the future.

\subsection{Other Resilient Network Technologies}

In addition to the advancements in 6G Mesh Network technology, various innovative technologies contribute to enhancing the resiliency of 3D wireless networks in disaster prevention and management scenarios. One notable example is the Virtualized Hybrid Satellite-Terrestrial Systems demonstrated by the VITAL (Virtualized hybrid satellite-TerrestriAl systems for resilient and fLexible future networks) project \cite{vital}, sponsored by the European Commission under the H2020 Research and Innovation program. By integrating software-defined network \cite{xiasurvey} and network function virtualization \cite{rakkiannan2023automated} into satellite networks, VITAL aims to provide federated resource management \cite{buyya2010federated} in hybrid satellite-terrestrial networks. This dynamic approach allows for virtualization and sharing of satellite communication platforms, 4G/5G backhauling services, and federated satellite-terrestrial access services, significantly contributing to the robustness and adaptability of the network.

Another technology enhancing the resiliency of 3D wireless networks is the Shared Access Terrestrial-Satellite Backhaul Network enabled by Smart Antennas (SANSA) \cite{sansa}. Designed to improve the performance of mobile wireless backhaul networks, SANSA proposes a terrestrial-satellite network to increase backhaul network capacity and resilience. Leveraging smart antennas, SANSA seamlessly integrates the satellite segment into terrestrial backhaul networks, offering adaptability to traffic demands, spectrum sharing, and resilience against link failures or congestion. This technology provides a flexible solution for efficiently routing mobile traffic, particularly beneficial in both low and highly populated areas. Moreover, SANSA's shared-access concept aligns with the need for efficient disaster prevention, such as large facility evacuation and timely reaction to wildfires, where a robust and adaptable network is crucial.

Furthermore, the integration of Edge Computing \cite{garcia2015edge} emerges as a pivotal technology to enhance the resiliency of 3D wireless networks. Edge Computing involves processing data closer to the source, reducing latency, and improving real-time decision-making capabilities. By decentralizing computational tasks, Edge Computing mitigates the impact of disruptions to centralized cloud servers, providing a more robust and responsive network. This technology is particularly valuable in disaster scenarios where split-second decisions are critical. The synergy between 3D wireless networks, Edge Computing, and technologies like SANSA ensures that critical information, including real-time sensor data and AI-driven analytics from large-scale evacuations to wildfire management, is processed swiftly, contributing to more proactive and resilient disaster prevention and management strategies.

\subsection{The Synergy Between NTN, 6G Mesh, and 3D Wireless Networks}
The relationship between 3GPP NTN, 6G mesh networks, and proposed 3D networks can lead to a more robust, efficient, and interconnected wireless communication ecosystem in the future. These concepts represent different facets of the ongoing evolution towards more integrated and pervasive communication systems, with each potentially complementing the others in future network architectures.

3GPP NTN focuses on extending cellular networks into non-terrestrial environments like satellites, providing global coverage and supporting various applications. 6G mesh networks are envisioned to be the backbone of future communication, leveraging advanced technologies like AI-based efficient and robust routing to enable ubiquitous connectivity. AI plays a critical role in all these paradigms by enabling intelligent network management, resource optimization, and dynamic adaptation to changing network conditions, thus acting as a key enabler for the efficient operation and evolution of these interconnected network architectures. Additionally, the proposed 3D networks aim to expand the concept of networks into the spatial dimension, potentially utilizing aerial and ground-based nodes to create a 3D network topology for enhanced capacity and coverage. 

In summary, 3GPP NTN can extend coverage to remote areas, while 6G mesh networks can enhance connectivity in urban environments, and 3D networks can provide multi-layered connectivity for increased reliability. Together, these technologies can offer enhanced capacity, throughput, and reduced latency, making them suitable for a wide range of applications including IoT, autonomous vehicles, and augmented reality. This combined approach can create a flexible and adaptable network that can dynamically adjust to user needs and environmental conditions, offering a seamless and reliable global connectivity, especially for managing and preventing disaster situations.

\begin{table*}  
\centering  
\caption{Real-world Prototypes}  
\label{tab:realworldprototypes}
\renewcommand{\arraystretch}{.5}
\scalebox{.8}{
\small
\begin{tabular}{|p{1.6cm}|p{4.9cm}|p{5.8cm}|p{4cm}|}  
\hline  
\textbf{Prototype} & \textbf{Prototype 1: Two-Tier UAV-based Low Power Wide Area Networks \cite{P1} } & \textbf{Prototype 2: A Direction of Arrival Estimation System for UAV-Assisted
Search and Rescue \cite{P2}} & \textbf{Prototype 3: An Intent-Based Reasoning System for Automatic Generation of Drone Missions for Public Protection and Disaster Relief \cite{P3}}\\  
\hline 
\textbf{Techniques Used} &   
\begin{itemize}  
\item Low Power Wide Area Network (LPWAN) system  
\item UAV base stations  
\item 3GPP NB-IoT  
\item LoRa  
\end{itemize} &   
\begin{itemize}  
\item Drone-based system for mobile phone localization  
\item Direction-of-arrival (DOA) estimation  
\item GSM based system  
\item MVDR beamformer algorithm for angle estimation  
\end{itemize} &   
\begin{itemize}  
\item Large language model Generative Pre-trained Transformer 3.5  
\item OpenAI's Whisper and Text-Davinci-003  
\item Parrot SDK  
\end{itemize} \\   
\hline  
\textbf{Hardware Involved} &   
\begin{itemize}  
\item UAVs  
\item Mobile base stations  
\item LPWAN user equipment  
\item NB-IoT module  
\end{itemize} &   
\begin{itemize}  
\item Antenna array  
\item Multichannel phase-coherent receivers  
\item Synchronization hardware for the BTS 
\item Server for data postprocessing and visualization  
\item Artix-7 FPGA  
\item FT600 USB3 SuperSpeed FIFO bridge  
\end{itemize} &   
\begin{itemize}  
\item Parrot ANAFI Ai Drone  
\item iPhone 11 Pro  
\item Server with Rust and Python  
\end{itemize} \\   
\hline  
\textbf{Use Case Analysis} &   
\begin{itemize}  
\item Suitable for deep rural environments without direct Tier 1 LPWAN network coverage  
\item Applications include agricultural, forestry, and environmental applications such as livestock or wild animal monitoring  
\end{itemize} &   
\begin{itemize}  
\item Post-disaster scenarios where search and rescue operations are needed  
\item The system targets mobile phones as indicators of buried victims, supplementing current methods with a drone-based multi-functional system  
\end{itemize} &   
\begin{itemize}  
\item Utilized for automatic generation of drone missions for public protection and disaster relief  
\end{itemize} \\   
\hline  
\textbf{Strengths} &   
\begin{itemize}  
\item Provides connectivity in remote areas  
\item Allows for mobile and static connectivity  
\item Can cover large areas  
\end{itemize} &   
\begin{itemize}  
\item Quick deployment after the SAR team arrives  
\item Efficient mapping and photographing of the stricken area  
\item Ability to capture hundreds of data points for DOA estimation  
\item Allows operations to happen in parallel with the traditional safety assessment of the rescue operation  
\end{itemize} &   
\begin{itemize}  
\item Provides an interface for drone control  
\item Allows for user intent to control drone operations  
\item Can generate complex drone missions  
\end{itemize} \\   
\hline  
\textbf{Key Innovation} &   
\begin{itemize}  
\item The two-tier system that combines a UAV-based mobile network with a macro-cellular network  
\end{itemize} &   
\begin{itemize}  
\item The use of a drone-based system to locate mobile phones as an indicator of buried victims, supplementing conventional search and rescue methods  
\end{itemize} &   
\begin{itemize}  
\item Use of AI for automatic generation of drone missions based on user intent  
\end{itemize} \\   
\hline  
\textbf{Limitations} &   
\begin{itemize}  
\item Depends on the availability of a Tier 1 macro-cellular network for backhaul links  
\item Deployment of UAVs as a capacity/coverage extension of the cellular network infrastructure is still under research and standardization  
\end{itemize} &   
\begin{itemize}  
\item The system is based on GSM, which isn't used everywhere and its global coverage will likely decline  
\item The signal reflections and attenuation in rubble piles can cause difficulties in data interpretation  
\end{itemize} &   
\begin{itemize}  
\item Limited to Parrot ANAFI AI drone  
\item Requires specific linguistic input for optimal performance  
\item Latency increases with complexity of commands  
\end{itemize} \\   
\hline  
\textbf{Potential Improvements} &   
\begin{itemize}  
\item Further optimization of LPWAN network parameters, like size of the Tier 2 cell, position and altitude of the Tier 2 drone base station, and the required capacity of the NB-IoT backhaul link  
\end{itemize} &   
\begin{itemize}  
\item More advanced algorithms could be used to improve the resolution of the spatial spectra  
\item Further development could explore the known properties of a GSM burst for significant improvements  
\end{itemize} &   
\begin{itemize}  
\item Improved accuracy for complex commands  
\item Reduced latency for complex commands 
\item Wider range of drone compatibility  
\end{itemize} \\   
\hline  
\end{tabular}} 
\label{table_1}
\end{table*}  

\section{The Role of AI in 3D Wireless Networks}
\label{sec:ai_3d_wireless} 
AI plays essential roles in advancing 3D wireless network development and operation through optimizing routing, managing resources, predicting maintenance needs, enabling autonomous operations, ensuring security, facilitating adaptive communication, and beyond. These functions empower autonomous decision-making, optimize resource usage, ensure security measures, and improve the overall functionality of satellite-based systems in preventing and managing disasters and emergency situations. In the context of 3D wireless networks for 6G networks, AI's potential extends to satellite communications, drone-connected networks, balloon-based systems, IoT \& remote sensing. It also plays a pivotal role in enhancing domains such as agriculture, logistics, maritime transportation, etc. This AI-powered framework is poised to serve as the foundation for the future of global telecommunications, offering advantages to remote, disconnected regions worldwide. With projections highlighting substantial areas on the planet lacking cellular connectivity, its adoption could notably improve conditions and detect abnormalities in these regions.

\subsection{AI for Network Optimization and Reliability}

AI algorithms play a crucial role in optimizing network operations within 3D wireless networks, ensuring robust and reliable communication under challenging conditions. Through predictive analytics, AI anticipates potential network disruptions due to atmospheric changes, satellite anomalies, or demand surges during crises. By adaptively reconfiguring network settings, AI maintains continuous communication, essential for effective disaster response \cite{Lei2021}.

\subsection{Enhanced Data Processing and Analysis}

In disaster management, swift data processing and analysis are critical. AI-driven 3D wireless networks efficiently manage large volumes of data from satellite imagery, aerial drones, and ground sensors, extracting real-time actionable insights. This accelerates decision-making, facilitating prompt evacuations, resource distribution, and emergency interventions \cite{Zhang2023}.

\subsection{Dynamic Resource Allocation}

AI technology enables dynamic resource allocation within 3D wireless networks \cite{Globecom2023Hualei}. Analyzing current network usage and predicting future demands allows AI algorithms to allocate bandwidth, power, and other resources efficiently. This optimizes network performance during terrestrial network impairments, reallocating resources to high-need areas during disasters.

\subsection{Autonomous Operations and Maintenance}

AI introduces self-healing and autonomous operational capabilities, enhancing 3D wireless network system resilience. AI detects, diagnoses, and autonomously corrects network issues, performing maintenance without human intervention. This minimizes downtime, ensuring 3D wireless networks remain operational during critical disaster periods \cite{Zhang2023}.

\subsection{Interference Management}

AI is crucial in managing and mitigating interference within 3D wireless networks. Continuous network monitoring for interference sources enables AI algorithms to adjust transmission parameters in real-time, minimizing disruptions. This ensures clear communication channels between satellites, UAVs, and ground stations, vital for coordinating disaster response efforts \cite{Warrier2023}.

\subsection{AI as Key Enabler for Mobile Edge Computing}

AI technology emerges as a pivotal enabler for Mobile Edge Computing (MEC) within Space-Air-Ground Integrated Networks (SAGIN), enhancing the computational efficiency and operational dynamism of mobile devices faced with stringent delay and energy constraints \cite{Globecom2023Bingqing}. Through AI, MEC systems within SAGIN can intelligently manage and optimize the deployment of computing, caching, and communication resources, ensuring low latency, high bandwidth, and extensive coverage to meet the robust demands of ubiquitous mobile applications. This integration emphasizes the profound transformation in quality of service delivery, pushing the boundaries of mobile application performance by leveraging the distributed intelligence and real-time processing capabilities of AI. Consequently, AI-underpinned MEC setups address the challenge of high dynamics, heterogeneity, and complex time-varying topology in SAGIN, marking a significant stride toward realizing the full potential of next-generation wireless networks \cite{OJCOMS1}.

\subsection{AI-Driven Mobile Edge Computing in 3D Wireless Networks}

AI significantly enhances MEC within 3D wireless networks, ensuring the seamless management and optimization of resources. Through strategic allocation of computing and communication resources, AI empowers MEC to fulfill the demanding performance requirements of various mobile applications, ensuring minimal latency and optimal efficiency. This integration of AI with MEC infrastructure propels the capabilities of 3D wireless networks to unprecedented levels, facilitating robust, adaptive network environments capable of supporting complex, resource-intensive applications across varied domains \cite{Qiu2022}.

\section{Combined Use Case Analysis}  
\label{sec:combined_use_case}
While terrestrial, UAV, and satellite communications can individually provide solutions for specific use cases, a holistic and more effective approach can be achieved by combining these technologies. This combined approach allows for comprehensive coverage and versatility, addressing the unique challenges presented in each use case more effectively. Table \ref{table_2} provides a detailed analysis of each combined scenario, outlining the sufficiency of each technology and describing the type of technology that will be employed.  

\begin{table*}
\centering
\caption{Combined Use Case Analysis}
\begin{tabular}{ |p{2cm}|p{3cm}|p{3cm}|p{3cm}|p{3cm}| }
\hline
\textbf{Use Case} & \textbf{Terrestrial Only} & \textbf{UAV Only} & \textbf{Satellite Only} & \textbf{3D Wireless Network} \\
\hline
Large Facility Evacuation & Limited coverage may fail to provide timely evacuation orders to all residents. Physical damage to networks can disrupt communication \cite{nguyen2023real}. & UAVs can relay evacuation orders to specific areas quickly but may not cover all affected areas due to limited operation time \cite{liu2023resource}. & Satellites can reach a wide area but the evacuation message might be delayed due to high latency \cite{esmat2023towards}. & The 3D network can provide reliable communication for timely evacuation orders to a wide area \cite{chen2022trajectory}.\\
\hline
Forecast and Alert Transmitter System & Limited coverage might not reach remote areas for timely alerts \cite{yao2023optimization}. & UAVs can quickly deploy to specific areas to broadcast alerts but are limited by operation time and coverage area \cite{ren2023novel}. & Satellites can broadcast to a wide area but alerts might be delayed due to high latency \cite{esmat2023towards}. & The 3D network can ensure timely and wide-area alerts, leveraging the strengths of all three systems \cite{chen2022trajectory}.\\
\hline
Transit Signal Priority in Emergency Evacuation & Terrestrial networks can provide localized traffic management but may be overwhelmed during emergencies \cite{yao2023optimization}. & UAVs can provide real-time traffic information but are limited by operation time and coverage area \cite{ren2023novel}. & Satellites can provide wide-area traffic information but it might be delayed due to high latency \cite{OJCOMS1}. & The 3D network can provide real-time, wide-area traffic management for efficient evacuation \cite{chen2022trajectory}.\\
\hline
Global Route Planning & Terrestrial networks can provide localized route planning but may not cover remote areas \cite{nguyen2023real}. & UAVs can provide real-time route updates but their operation time and coverage area are limited \cite{liu2023resource}. & Satellites can provide wide-area route planning but their high latency can result in delayed updates \cite{OJCOMS1}. & The 3D network can provide real-time, wide-area route planning for efficient evacuation \cite{chen2022trajectory}.\\
\hline
Life Detection & Terrestrial networks can provide localized life detection but may not cover remote areas \cite{nguyen2023real}. & UAVs can provide real-time life detection information but their operation time and coverage area are limited \cite{liu2023resource}. & Satellites can provide wide-area life detection but their high latency can result in delayed detection \cite{OJCOMS1}. & The 3D network can provide real-time, wide-area life detection for efficient rescue operations \cite{chen2022trajectory}.\\
\hline
Wi-Fi Indoor Localization and Navigation & Terrestrial networks can provide localized indoor localization but may not cover all areas due to physical obstructions \cite{nguyen2023real}. & UAVs can provide real-time indoor navigation information but their operation time and coverage area are limited \cite{liu2023resource}. & Satellites may not provide accurate indoor localization due to signal obstructions \cite{esmat2023towards}. & The 3D network can provide accurate and real-time indoor localization and navigation \cite{chen2022trajectory}.\\
\hline
Wildfire Management & Terrestrial networks can provide localized wildfire monitoring but may not cover remote forest areas \cite{nguyen2023real}. & UAVs can provide real-time wildfire tracking but their operation time and coverage area are limited \cite{liu2023resource}. & Satellites can provide wide-area wildfire monitoring but their high latency can result in delayed alerts \cite{OJCOMS1}. & The 3D network can provide real-time, wide-area wildfire monitoring and alerts, ensuring timely firefighting operations \cite{chen2022trajectory}.\\
\hline
\end{tabular}
\label{table_2}
\end{table*}
In all these use cases, the combined 3D wireless network, integrating terrestrial, UAV, and satellite networks, provides a solution that leverages the strengths of each individual technology to address their limitations.

\section{Challenges and Future Directions}
\label{sec:challenges_future_directions}
\begin{figure}[t]
	\centering
\includegraphics[width=\linewidth]{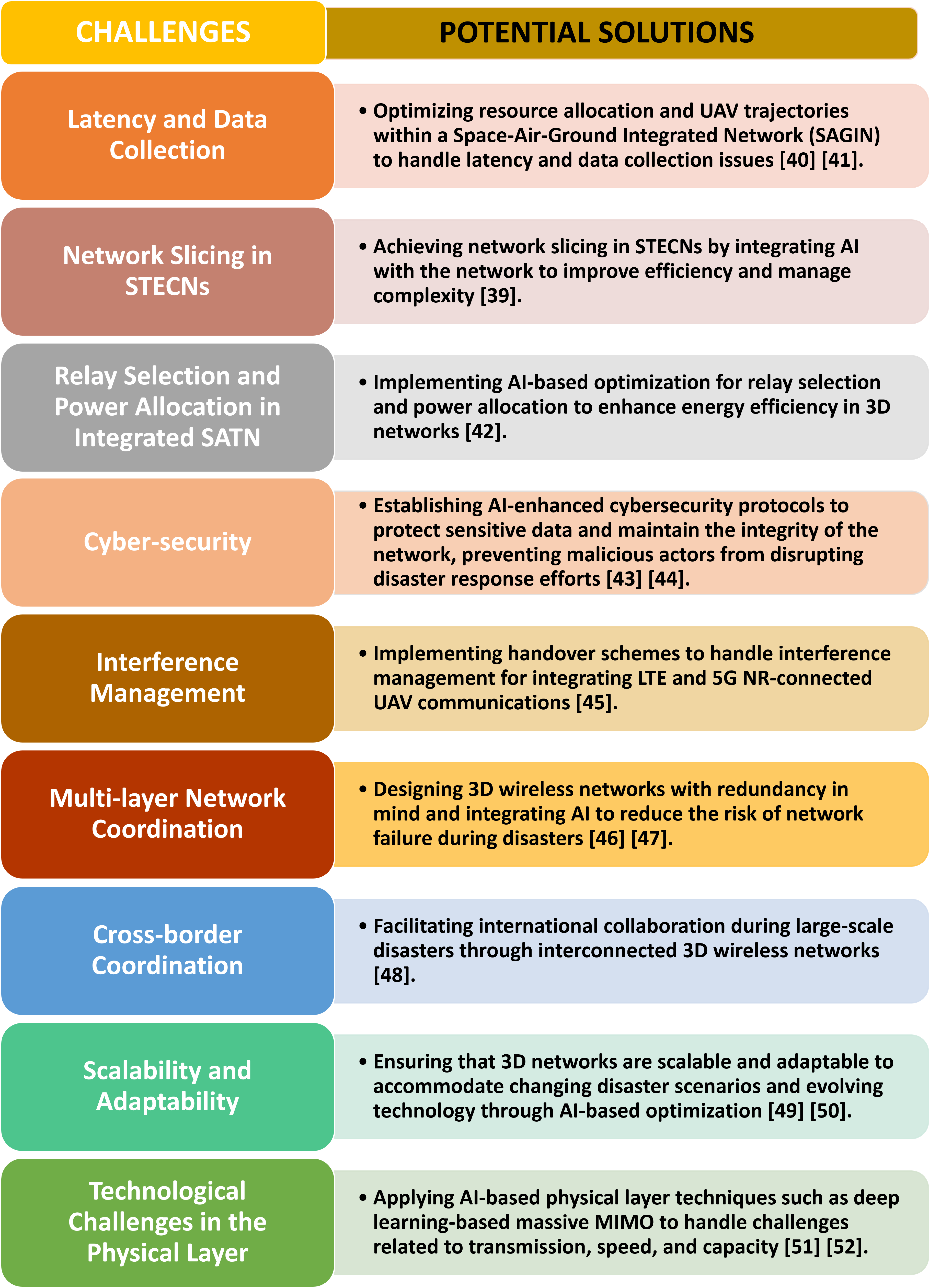}
	\caption{Challenges and Potential Solutions in 3D Wireless Networks.}		
 \label{fig_6}
\end{figure}

The 3D wireless network is an emerging area of research and development with promising potential through the usage of more advanced 5G features and standards or possibly the transition to 6G deployment or futuristic smart cities.

The mobile operators are no longer restricted to the terrestrial domain. Instead, a consortium of terrestrial telecommunication operators, satellite telecommunication companies, regulators, air traffic control centers, and other government entities shall be involved in order to have a smooth and streamlined integration. This integration of 3D wireless networks into various sectors has the potential to reshape industries and improve the quality of life in many ways. This will provide flexible and versatile services to cover numerous critical applications and scenarios, some of which are discussed in this paper.

Despite impressive advancements in 3D wireless networks, certain persisting challenges invite focused research efforts. Jung et al., for instance, underscored certain obstacles related to latency and data collection while discussing the role of Marine IoT systems within 6G networks \cite{jung2023marine}. Likewise, Hu et al. studied how best to optimize resource allocation and UAV trajectories within a SAGIN, calling attention to the challenges therein \cite{hu2023joint}. In view of the increasing demand for wide coverage and strict adherence to administrative domains, Esmat et al. highlighted the complexity in achieving Network Slicing in Satellite-Terrestrial Edge Computing Networks (STECNs)  \cite{esmat2023towards}. Similarly, Shi et al. research on multi-UAV relay systems underscored the importance of optimized relay selection and power allocation for enhancing energy efficiency in integrated Satellite-Aerial-Terrestrial Networks (SATN) \cite{shi2022energy}.

Collectively, the research landscape in this field continues to identify potential areas for ameliorating 3D wireless networks' performance, while also shining a light on challenges that warrant dedicated scientific inquiry to find effective solutions.
As depicted in Fig. \ref{fig_6}, these challenges span a wide range of areas, from latency and data collection to cybersecurity and interference management, among others. The Fig. \ref{fig_6} also elucidates potential solutions to these challenges, many of which leverage advanced AI-based optimization schemes, conventional optimization strategies, and handover schemes. By systematically addressing these challenges, we can unlock the full potential of 3D wireless networks and greatly enhance our disaster management capabilities.
\nocite{jung2023marine,hu2023joint,esmat2023towards,shi2022energy,AI1,AI7,H1,AI3,AI9,AI5,AI2,AI6,AI4,AI8}

\section{Conclusion}
\label{sec:conclusion}
In conclusion, the transformative potential of 3D wireless networks in disaster prevention and management is undeniable. By integrating terrestrial, aerial, and satellite technologies, 3D networks enhance communication resilience, real-time situational awareness, and efficient resource allocation during crises. This paper has explored a wide range of use cases, from large facility evacuations to wildfire management, highlighting the versatility and practical feasibility of these networks.  This is further underscored by the success of various real-world prototypes which have showcased the practical feasibility and operational efficacy of these networks in diverse scenarios. However, it is crucial to note that there are challenges posed by aspects like cybersecurity, cross-border coordination, and physical layer technological hurdles that require further research and development. As technology continues to evolve, the development of 3D wireless networks and their integration into disaster management strategies will be instrumental in building safer, more resilient communities.
\printbibliography

\end{document}